\documentclass[12pt, letterpaper] {article}
\usepackage[utf8]{inputenc}
\usepackage{amsmath}
\usepackage{mathtools}
\usepackage{amsfonts,amssymb,amsthm}
\usepackage{graphicx} 
\graphicspath{{./Pics/}}
\usepackage[height=1in, width=0in, left=1in, right=1in, top=1in, bottom=1in]{geometry}
\usepackage{setspace}
\usepackage[symbol]{footmisc}

\begin{document}
\begin{center}
    \Large{\textbf{Eigen Value Statistics of Long-Term Monthly Average Temperature of Meghalaya, India}} \\[20pt]
    \large{\textbf{Raju Kalita\footnote[1]{kalita.raju.nehu@gmail.com}}, and \textbf{Atul Saxena}} \\[1pt]
    \small{Department of Physics, North-Eastern Hill University, Shillong-22, India} \\[40pt]
\end{center} 
\begin{abstract}
    We use Random Matrix Theory (RMT) to describe the eigenvalue spacing of Meghalaya's historical monthly average temperature ($T_{avg}$) in grids. For that, the Nearest Neighbor Spacings ($S_i$) of the eigenvalues of the correlation matrices were found out for 1428 consecutive eigenvalue pair differences. It is found that the distribution of $S_i$ follows Brody distribution at a correlation value of $\beta=0.045$. This value of $\beta(0.045)$ indicates weak repulsion among the eigenvalues as it is closer to Poisson fluctuations, meaning there is a weak correlation among the grids. 
\end{abstract}

\begin{spacing}{1.5}
\section{Introduction}
The theory of the Random Matrix is quite successful in understanding the amount of correlation in different time series. It was Eugene P. Wigner who first applied the technique of random matrix theory to model the nuclei of heavy atoms \cite{a1}. Since then, it has been used remarkably in many multivariate data sets like financial \cite{b1}, human electroencephalographic \cite{c1}, city transport \cite{krbalek2000statistical}, internet traffic \cite{barthelemy2002large}, atmospheric data \cite{santhanam2001statistics}, sea surface temperature \cite{santos2020global}, etc. The statistical properties of random matrix ensembles such as Gaussian Orthogonal (GOE), Gaussian Unitary (GUE), and Gaussian Symplectic (GSE) have been studied extensively by pioneers like Wigner, Dyson, Mehta, etc. \cite{mehta2004random}. The main advantage of this theory is that it can correctly describe the spectral statistics of various complex, chaotic systems \cite{guhr1998random}.

Moreover, the spectral properties of the correlation matrices arising from the random matrix can separate signals from noise. The short-range correlations are mainly observed by studying the Nearest Neighbour Spacing Distributions (NNSD) of eigenvalues arising from the correlation matrices \cite{nieminen2017random}. Since the NNSD of eigenvalues of the correlation matrices gives the nature of correlation, using RMT, their different modes of randomness can be predicted. 

This paper shows that the empirical correlation matrices arising from the half-degree latitude-longitude $T_{avg}$ grids over Meghalaya can be modeled as random matrices chosen from an appropriate ensemble.

\section{Study area and data used}
The area under study covers almost the entire state of Meghalaya, located in the North-Eastern part of India (Fig. 1(a)). The hilly terrain of Meghalaya mainly comprises of three mainlands; Khasi Hills (central region), Jaintia Hills (eastern part), and Garo Hills (western part). It lies in-between $25.00^0N$ to $26.10^0N$ latitude and $89.45^0E$ to $92.45^0E$ longitude covering an area of 22,549 square kms \cite{haridasan1985forest} (Fig. 1(b)). 

\begin{figure} [h]
    \centering
    \includegraphics[width=12cm, height=6cm]{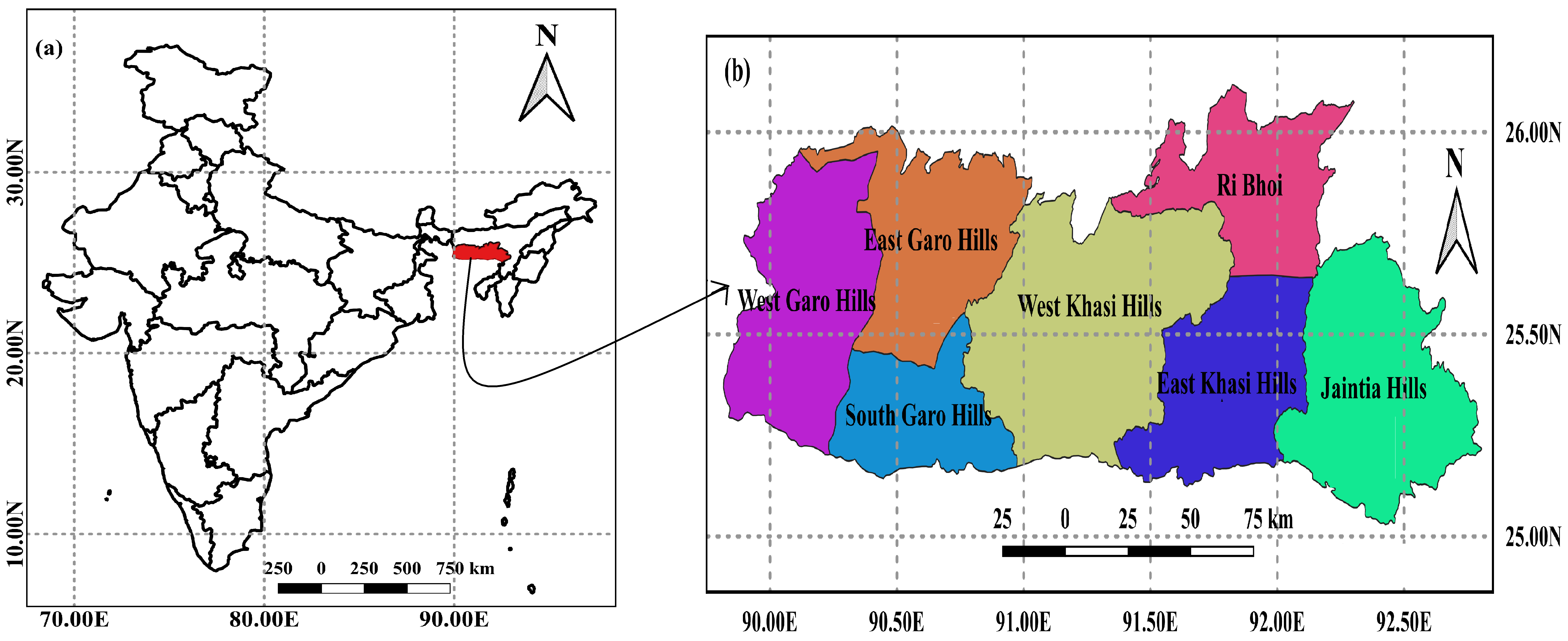}
    \caption{Location of study area (a) India and (b) Meghalaya in grids ($0.5^0\times 0.5^0$).}
\end{figure}

The data set for monthly average temperature has been extracted from $0.5^0\times 0.5^0$ latitude-longitude grid boxes of CRU TS 4.04 over Meghalaya \cite{harris2020version} using the Google Earth interface. Grids are sorted from left top to right bottom in a logical sequence (Fig. 2). Data set for 10 out of 11 grids from 1901 to 2019 were arranged in a matrix form in such a way that the first matrix for January 1901 has five values (grid no 1 to 5) in one row (center latitude: $25.75^0N$; center longitude: $90.25^0E$, $90.75^0E$, $91.25^0E$, $91.75^0E$, $92.25^0E$) and the rest five values (grid no 6 to 10) in the second row (center latitude: $25.25^0N$; center longitude: $90.25^0E$, $90.75^0E$, $91.25^0E$, $91.75^0E$, $92.25^0E$).   

\begin{figure} [h]
    \centering
    \includegraphics[width=12cm, height=8cm]{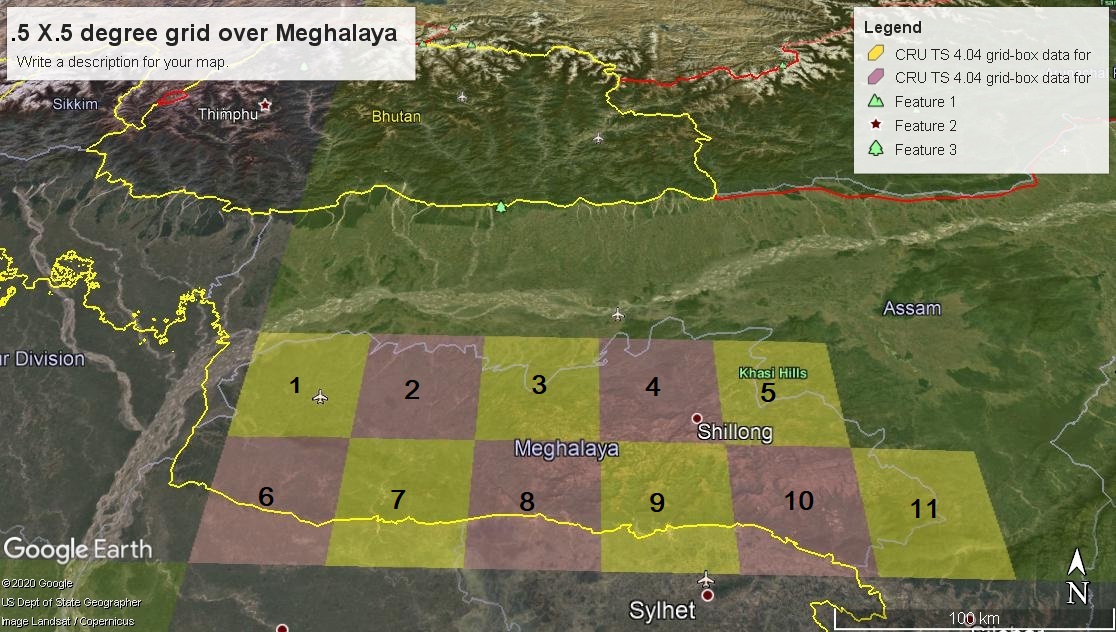}
    \caption{Google Earth image of CRU TS v4.04 in half-degree grids over Meghalaya numbered as 1 to 11 from left top to right bottom.}
\end{figure}

\section{Construction and evaluation of random matrices}

The RMT framework defines the grid system as an ensemble matrix $W_{2\times 5}$ with random inputs. This random matrix W contains each month's data of 10-time series  $X_j(k)$ where $j = 1,2,...,10$ (grid position) and $k=1,2,...,1428$ (no. of months in ascending order). Since there are $1428$ months from January 1901 to December 2019, each random matrix $W$ corresponds to a particular month of each year. Then each of the correlation matrix $C_{2\times 2}$ is constructed from the multivariate random matrix W of two rows and five columns given by,

\begin{equation}
 C_{ij}=\frac{1}{5}\sum_{k=1}^{5}{x_i(k)x_j(k)} 
\end{equation}
Where $x_i(k)$ corresponds to the transpose of matrix $W$, and $x_j(k)$ corresponds to matrix $W$. With $\lambda_i$, the eigenvalues, and $\vec{\nu_l}$, the eigenvectors, the correlation matrix is,

\begin{equation}
    C\vec{\nu_l}=\lambda_i\vec{\nu_l} 
\end{equation}

The largest eigenvalue of each correlation matrix is then sorted as $\lambda_1 \leq \lambda_2 \leq \lambda_3....... \leq \lambda_{1428}$, with their increasing size. Now the distribution of these eigenvalues is closely related to the amount of correlation in the random inputs of the multivariate data set \cite{muller2005detection}. The Nearest Neighbor Spacings $S_i$ were then found out as
\begin{equation}
   S_i=\frac{\lambda_{i+1}-\lambda_i}{<\lambda_{i+1}-\lambda_i>}
\end{equation} 
where $i=1,2,...,1427$ and $<\lambda_{i+1}-\lambda_i>$ denotes average value over 1428 consecutive eigenvalue pair differences. Studies have shown that the probability distribution is well described by Brody distribution \cite{brody1973statistical}. 
\begin{equation}
    P(S_i)=\left[\Gamma\left(\frac{2+\beta}{1+\beta}\right)\right]^{(1+\beta)}{(1+\beta)} S_i^{\beta} e^{-\left[\Gamma\left(\frac{2+\beta}{1+\beta}\right)\right]^{(1+\beta)}S_i^{(1+\beta)}}     
\end{equation}
Where $\Gamma(x)$ is the Gamma function. The parameter $\beta$ in the above distribution classifies the correlation in the system with respect to its probability distribution. When there is no correlation, the spacing of levels is very close and $\beta \to 0$ and leads to Poisson distribution given by,
\begin{equation}
    P(S_i)=e^{-S_i} 
\end{equation}
However, when a correlation is present, then the level repels each other and $\beta \to 1$, and this leads to GOE fluctuations given by, 
\begin{equation}
    P(S_i)=\frac{\pi}{2}S_ie^{-\frac{\pi}{4}S_i^2}  
\end{equation}
This Poisson to GOE fluctuation gives the measure of correlation in the system of the multivariate data set \cite{sakhr2005poisson}.
\section{Result and discussion}

After extracting the Eigenvalues from the random correlation matrices $C_{ij}$, their distribution is plotted analytically with a non-parametric fitting (Fig. 3). It is observed that most of the eigenvalues lie on the higher side. This indicates uniformity in the next-to-next eigenvalue, as a result of which the eigenvalues are likely to reside close to each other.

\begin{figure} [ht]
    \centering
    \includegraphics[width=11cm, height=7cm]{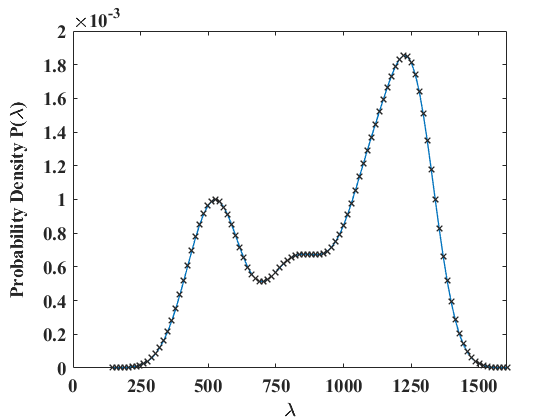}
    \caption{Probability density of eigenvalues $(\lambda_i)$ of the correlation matrices.}
\end{figure}

To find the Nearest Neighbour Spacing Distribution (NNSD), we plot the non-parametric histogram fitting of $S_i$ (Fig. 4 [blue line]). After that, the best fit is adjusted using equation (4) and is obtained at the Brody parameter, $\beta=0.045$. This value of $\beta$ indicates a fluctuation near to Poisson distribution. This means that though the level spacing repulsion is very small, it shows a very weak correlation among half-degree temperature grids of Meghalaya.

\begin{figure} [ht]
    \centering
    \includegraphics[width=11cm, height=7cm]{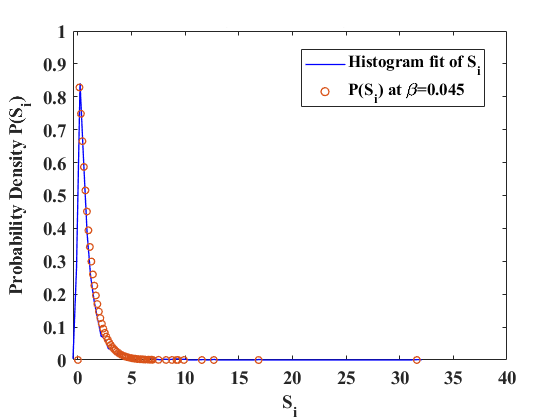}
    \caption{Nearest Neighbour Spacing Distribution $(S_i)$ of eigenvalues $(\lambda_i)$ of the correlation matrices.}
\end{figure}
\begin{figure} [ht]
    \centering
    \includegraphics[width=11cm, height=7cm]{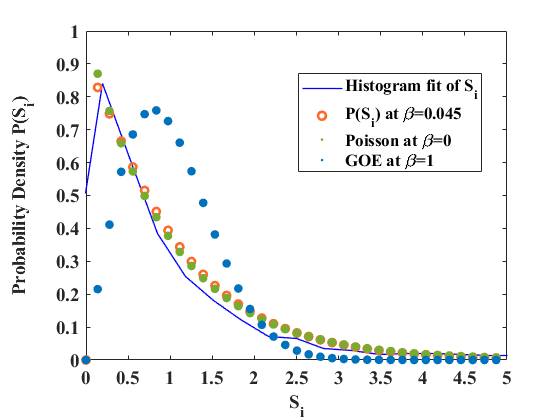}
    \caption{Comparison of fitted Nearest Neighbour Spacing Distribution $(S_i)$ of eigenvalues $(\lambda_i)$ 	with Poisson (light blue dots) and GOE (light green dots) fluctuations.}
\end{figure}
The 119 years for CRU TS v4.04 Tavg data analysis in RMT frameworks reveals that the half-degree grids over Meghalaya are weakly correlated. The NNSD shows fluctuations closer to Poisson than the GOE ensemble (Fig. 5). Thus, in the present work, we could replace the analytical spacing distribution with an ensemble of random matrices that follows Brody distribution at $\beta=0.045$, which indicates a weak random fluctuation in the average temperature that existed over the Meghalaya throughout the period 1901 to 2019.

\end{spacing}
\newpage

\bibliographystyle{ieeetr}
\bibliography{./bibliography.bib}

\begin{thebibliography}{10}

\bibitem{a1}
E.~P. Wigner, ``Random matrices in physics,'' {\em SIAM review}, vol.~9, no.~1,
  pp.~1--23, 1967.

\bibitem{b1}
V.~Plerou, P.~Gopikrishnan, B.~Rosenow, L.~A.~N. Amaral, T.~Guhr, and H.~E.
  Stanley, ``Random matrix approach to cross correlations in financial data,''
  {\em Physical Review E}, vol.~65, no.~6, p.~066126, 2002.

\bibitem{c1}
P.~{\v{S}}eba, ``Random matrix analysis of human eeg data,'' {\em Physical
  review letters}, vol.~91, no.~19, p.~198104, 2003.

\bibitem{krbalek2000statistical}
M.~Krb{\'a}lek and P.~Seba, ``The statistical properties of the city transport
  in cuernavaca (mexico) and random matrix ensembles,'' {\em Journal of Physics
  A: Mathematical and General}, vol.~33, no.~26, p.~L229, 2000.

\bibitem{barthelemy2002large}
M.~Barth{\'e}lemy, B.~Gondran, and E.~Guichard, ``Large scale
  cross-correlations in internet traffic,'' {\em Physical Review E}, vol.~66,
  no.~5, p.~056110, 2002.

\bibitem{santhanam2001statistics}
M.~Santhanam and P.~K. Patra, ``Statistics of atmospheric correlations,'' {\em
  Physical Review E}, vol.~64, no.~1, p.~016102, 2001.

\bibitem{santos2020global}
E.~F. Santos, A.~L. Barbosa, and P.~J. Duarte-Neto, ``Global correlation matrix
  spectra of the surface temperature of the oceans from random matrix theory to
  poisson fluctuations,'' {\em Physics Letters A}, vol.~384, no.~27, p.~126689,
  2020.

\bibitem{mehta2004random}
M.~L. Mehta, {\em Random matrices}.
\newblock Elsevier, 2004.

\bibitem{guhr1998random}
T.~Guhr, A.~M{\"u}ller-Groeling, and H.~A. Weidenm{\"u}ller, ``Random-matrix
  theories in quantum physics: common concepts,'' {\em Physics Reports},
  vol.~299, no.~4-6, pp.~189--425, 1998.

\bibitem{nieminen2017random}
J.~M. Nieminen and L.~Muche, ``A random matrix model whose eigenvalue spacings
  are closely described by the brody distribution.,'' {\em Acta Physica
  Polonica B}, vol.~48, no.~4, 2017.

\bibitem{haridasan1985forest}
K.~Haridasan and R.~R. Rao, {\em Forest flora of Meghalaya}.
\newblock Dehra Dun India, 1985.

\bibitem{harris2020version}
I.~Harris, T.~J. Osborn, P.~Jones, and D.~Lister, ``Version 4 of the cru ts
  monthly high-resolution gridded multivariate climate dataset,'' {\em
  Scientific data}, vol.~7, no.~1, p.~109, 2020.

\bibitem{muller2005detection}
M.~M{\"u}ller, G.~Baier, A.~Galka, U.~Stephani, and H.~Muhle, ``Detection and
  characterization of changes of the correlation structure in multivariate time
  series,'' {\em Physical Review E}, vol.~71, no.~4, p.~046116, 2005.

\bibitem{brody1973statistical}
T.~Brody, ``A statistical measure for the repulsion of energy levels,'' {\em
  Lettere al Nuovo Cimento (1971-1985)}, vol.~7, no.~12, pp.~482--484, 1973.

\bibitem{sakhr2005poisson}
J.~Sakhr and J.~M. Nieminen, ``Poisson-to-wigner crossover transition in the
  nearest-neighbor statistics of random points on fractals,'' {\em Physical
  Review E}, vol.~72, no.~4, p.~045204, 2005.

\end{thebibliography}

\end{document}